\documentclass[a4paper, amsfonts, amssymb, amsmath, reprint, twoside]{revtex4-2}

\usepackage{graphicx}
\usepackage{dcolumn}
\usepackage{bm}
\usepackage{multirow} 
\usepackage{caption}
\usepackage{amsmath}
\usepackage{hyperref}

\usepackage{scalerel}
\usepackage{tikz}
\usetikzlibrary{svg.path}

\definecolor{orcidlogocol}{HTML}{A6CE39}
\tikzset{
  orcidlogo/.pic={
    \fill[orcidlogocol] svg{M256,128c0,70.7-57.3,128-128,128C57.3,256,0,198.7,0,128C0,57.3,57.3,0,128,0C198.7,0,256,57.3,256,128z};
    \fill[white] svg{M86.3,186.2H70.9V79.1h15.4v48.4V186.2z}
                 svg{M108.9,79.1h41.6c39.6,0,57,28.3,57,53.6c0,27.5-21.5,53.6-56.8,53.6h-41.8V79.1z M124.3,172.4h24.5c34.9,0,42.9-26.5,42.9-39.7c0-21.5-13.7-39.7-43.7-39.7h-23.7V172.4z}
                 svg{M88.7,56.8c0,5.5-4.5,10.1-10.1,10.1c-5.6,0-10.1-4.6-10.1-10.1c0-5.6,4.5-10.1,10.1-10.1C84.2,46.7,88.7,51.3,88.7,56.8z};
  }
}

\newcommand\orcidicon[1]{\href{https://orcid.org/#1}{\mbox{\scalerel*{
\begin{tikzpicture}[yscale=-1,transform shape]
\pic{orcidlogo};
\end{tikzpicture}
}{|}}}}

\begin{document}

\preprint{APS/123-QED}

\title{Thermodynamics of Geodetic Brane Gravity}

\author{Gilberto Aguilar-Pérez\orcidicon{0000-0001-6821-4564}}
\email{gilaguilar@uv.mx}

\author{Giovany Cruz\orcidicon{0000-0002-2877-6922}}
\email{giocruz@uv.mx }\thanks{(Corresponding author)}

\author{Miguel Cruz\orcidicon{0000-0003-3826-1321}}
\email{miguelcruz02@uv.mx}

\author{Efraín Rojas\orcidicon{0000-0002-6898-2720}}
\email{efrojas@uv.mx}
\affiliation{Facultad de F\'{\i}sica, Universidad Veracruzana 91097, Xalapa, Veracruz, M\'exico.}

\date{\today}

\begin{abstract}
In this work, we explore the effect at the cosmological level of the extra contribution arising from the Geodetic Brane Gravity model within a thermodynamical perspective. As already known, the universe seen as an extended object embedded within a higher dimensional spacetime modifies the dynamical background equations, which in turn results in correction contributions to the entropy and temperature of the apparent horizon. Additionally, we investigate the possibility that the apparent horizon and the bulk remain in thermal equilibrium across various matter contents, demonstrating that such properties are highly sensitive to the equation-of-state parameter. 
\end{abstract}

\maketitle


\section{\label{sec1}Introduction}

The seminal contributions of Bekenstein and Hawking \cite{bekenstein2020black, hawking1975particle} established that the event horizon of black holes can be endowed with thermodynamic attributes---notably temperature and entropy---thereby revealing an unexpected link between gravitation and thermodynamics. Building on this insight, Jacobson demonstrated that Einstein's equations may be interpreted as an equation of state \cite{jacobson1995thermodynamics} since it can be obtained from the relation $\delta Q = TdS$, which is a fundamental relation connecting heat, entropy and temperature. This is the well-known Clausius relation of classical thermodynamics \cite{callen}. 
These findings indicate that gravity inherently possesses a thermodynamic description, rather than just showing superficial similarities to thermodynamic systems.\\

This approach has been applied within the cosmological context, indicating that the apparent horizon of the universe functions similarly to the event horizon of a black hole \cite{Hayward_1998, bak2000cosmic}. Consequently, the Friedmann equations are found to be equivalent to the first law of thermodynamics at the apparent horizon \cite{cai2005first}, which allows the exploration of the universe through the temperature and entropy linked to this horizon. Within this line of research, two routes can be identified. The first one consists of starting from cosmological equations of the Friedmann type, whether arising from standard general relativity or from modified gravity theories, and analyzing them from a thermodynamic perspective in order to extract properties such as temperature and entropy. The second follows an inverse path, that is, one postulates an entropy different from the standard one motivated, for instance, by quantum corrections or by alternative formulations of gravity, and from this assumption, reconstructs the effective cosmological equations governing the dynamics of the universe \cite{sheykhi2010entropic}.\\

In this work, we focus on the first route, applying it in the framework of brane-like universes, where similar thermodynamic treatments of the apparent horizon have also been investigated
\cite{sheykhi2007thermodynamical, sheykhi2007deep, sheykhi2009thermodynamical}. As is already known, braneworld models in cosmology stem primarily from the need to reconcile the standard model of particle physics and general relativity; such compatibility can be achieved with the introduction of extra spatial dimensions. These models, inspired by string theory, propose that our observable universe (composed of 3 spatial dimensions and 1 temporal) is a lower-dimensional surface, or {\it brane}, embedded within a higher-dimensional spacetime. This framework offers potential solutions to the hierarchy problem (the large disparity between the gravitational and electroweak scales) by allowing gravity to propagate through the higher-dimensional spacetime, thereby weakening its apparent strength on the brane. 

Furthermore, some braneworld scenarios, such as the Randall-Sundrum models, naturally generate modifications to general relativity at high energies, which could potentially account for the accelerating expansion of the universe without requiring a cosmological constant or dark energy contribution \cite{Shiromizu_2000, Randall_1999}. In contrast to the aforementioned model, there exist modifications to general relativity at low energies, a well-known example of this is the Dvali-Gabadadze-Porrati (DGP) model \cite{Dvali_2000}, both approaches are needed since at low energies, the zero-mode of the graviton is expected to dominate on the brane. On the other hand, at high energies the massive modes of the graviton dominate over the zero-mode, and the gravity on the brane behaves increasingly 5-dimensional. Due to its relevance, the cosmological formulation of the DGP model was carried out in \cite{Deffayet_2001}. For a complete review on brane-gravity and cosmology, see~\cite{Maartens_2010}.\\ 

In particular, we consider the \textit{Geodetic Brane Gravity} (GBG) theory, based on the Regge-Teitelboim model (RT), 
\cite{regge2016general}, which represents a minimal extension of general 
relativity, and describes the universe as an extended object embedded floating
{\it geodesically} in a flat background spacetime. The GBG, also 
known as \textit{embedding gravity}, has generated significant research interest as 
an engaging formulation to describe gravity, its cosmological implications in 
extra dimensions, as a suitable alternative for the quantization of gravity, besides allowing 
the introduction of additional fictitious matter, also labeled as dark matter,~\cite{deser1976new, pavvsivc1985classical, tapia1989gravitation, davidson1998quantum, davidson2003geodetic, paston2007canonical, cordero2009ostrogradski,estabrook2010hilbert,banerjee2014new, capovilla2022ostrogradsky, fabi2022cosmic}. In contrast to general relativity, where the fundamental degrees of freedom are given by the components of the metric tensor, this formulation introduces the physical degrees of freedom through embedding functions that describe how the brane is immersed in the ambient space. This scheme results in dynamical equations that are more comprehensive than those in general relativity, so that each solution to Einstein's equations is encompassed within this framework. Furthermore, within the cosmological scenario, an extra parameter in the Friedmann equations arises, and it is linked to the brane energy, influencing how much the cosmology in this model deviates from the conventional one grounded in general relativity \cite{davidson1999cosmology}. Indeed, this is the fingerprint of the extra dimension, as well as the backbone, which allows us to explore its impact on the thermodynamical properties of the universe within the framework of GBG cosmology. Considering the robustness we acknowledge in general relativity, for our approach we will assume that only small deviations from general relativity are permissible. We recognize that the original RT framework encounters theoretical difficulties, particularly the ambiguity in choosing the embedding and in specifying the background geometry \cite{deser1976new}. However, in contemporary settings, such embedding approaches are fruitfully interpreted as effective field theories, where the extrinsic curvature serves as an effective description of dark sector dynamics, \cite{Paston_2012, Paston_2023, Rojas2024dark}. In this work, we adopt this effective viewpoint to investigate how these geometric degrees of freedom affect the thermodynamic properties of the universe.


This work is organized as follows: in Section \ref{sec:apparent} we comment on the general properties associated with the apparent horizon and its corresponding physical quantities. The GBG model is introduced in Section \ref{sec:gbg} where we focus on its dynamical equations and the energy parameter associated with the brane. In Section \ref{sec:thermo} we analyze the thermodynamic formulation of the model and compute the correction contributions to the entropy of the apparent horizon. In addition, we discuss the behavior of the heat capacities for a specific matter content. We corroborate that the approximation used in our development is viable to describe the late times evolution of the universe. In Section \ref{sec:final} we give our perspectives on this formulation and the final comments of the work.

\section{Apparent horizon and thermodynamics}
\label{sec:apparent}

We consider a Friedmann-Lemaître-Robertson-Walker (FLRW) spacetime written in the spherically symmetric form
	\begin{equation}
		ds^{2}=h_{ab}\,dx^{a}dx^{b}+R^{2}(t,r)\,d\Omega_{2}^{2},
	\end{equation}
where the indices $a,b=0,1$ correspond to the temporal and radial coordinates $(t,r)$, and
	\begin{equation}
		R(t,r)=a(t)\,r
	\end{equation}
is the areal radius. The two--dimensional metric that governs the $(t,r)$ submanifold is
	\begin{equation*}
		h_{ab}=\mathrm{diag}\!\left(-1,\;\frac{a^{2}(t)}{1-k r^{2}}\right),
		\,\,\, 
		h^{ab}=\mathrm{diag}\!\left(-1,\;\frac{1-k r^{2}}{a^{2}(t)}\right),
	\end{equation*}
while $d\Omega_{2}^{2}$ is the metric of the unit two--sphere. The apparent horizon is defined as the surface on which the expansion of outward future-directed null geodesics vanishes \cite{Faraoni}. This geometric requirement is encoded in the condition
	\begin{equation}
		h^{ab}\,\partial_{a}R\,\partial_{b}R=0.
	\end{equation}
Using $\partial_{t}R=\dot{a}\,r=HR$ and $\partial_{r}R=a$, the horizon condition becomes $	- (H R)^{2} + \frac{1-k r^{2}}{a^{2}}\,a^{2} = 0$	or, equivalently, $H^{2}R^{2} = 1 - k\,\frac{R^{2}}{a^{2}}$. Solving for $R$ gives the radius of the apparent horizon,
	\begin{equation}
		R_{A}=\frac{1}{\sqrt{H^{2}+k/a^{2}}}.
	\end{equation}
This surface plays a key role in dynamical spacetimes. Unlike the event horizon, which requires knowledge of the full future causal structure, the apparent horizon is defined locally and responds instantaneously to the cosmic expansion. In FLRW geometry, it separates regions where outgoing null rays expand from those where they begin to converge, thereby acting as a \emph{causal horizon} for spacetime \cite{sheykhi2007thermodynamical}. Physical signals originating beyond $R_A$ cannot propagate outward to influence an observer inside this radius, making the apparent horizon the physically relevant causal boundary for thermodynamic analyzes of an evolving universe \cite{bak2000cosmic}.\\
	
The thermodynamic character of this surface emerges from the fact that the gravitational dynamics in the FLRW spacetime can be rewritten in a form reminiscent of the first law of thermodynamics \cite{cai2005first}. The apparent horizon is endowed with an entropy proportional to its area, reducing to the Bekenstein--Hawking expression in general relativity and with a temperature determined by its \emph{surface gravity}. For spherically symmetric, dynamical spacetimes, the surface gravity may be written in the form
	\begin{equation}
		\kappa=\frac{1}{\sqrt{-h}}\partial_a\!\left(\sqrt{-h}\,h^{ab}\partial_b R\right),
	\end{equation}
which evaluates at the apparent horizon $R=R_A$ to
	\begin{equation}\label{superfi}
		\kappa=-\frac{1}{R_A}\left(1-\frac{\dot R_A}{2 H R_A}\right).
	\end{equation}
In the quasi-static regime, $\dot R_A\ll H R_A$, one recovers $\kappa\simeq -1/R_A$, showing that the horizon behaves locally like a Schwarzschild sphere of radius $R_A$.
	
The temperature associated with the apparent horizon is then defined as
	\begin{equation}\label{eq:Tkappa}
		T=\frac{\kappa}{2\pi}.
	\end{equation}
Cosmological applications typically use $T=|\kappa|/2\pi$, since $\kappa<0$ in an expanding universe. This notion of temperature corresponds to that measured by an observer at the horizon and is crucial for expressing the gravitational field equations in thermodynamic form.
	
These considerations naturally lead to the unified first law
	\begin{equation}
		dE = T\,dS + W\,dV,
	\end{equation}
where $E$ is the generalized Misner--Sharp energy contained within the apparent horizon \cite{misner1964relativistic}, and \[W=\tfrac{1}{2}(\rho-p)\] is the work density of the cosmological fluid. In modified gravity theories, the entropy associated with the apparent horizon cannot be assumed beforehand; instead, it must be defined so that the first law reproduces the modified Friedmann equations \cite{akbar2007thermodynamic,eling2006non,cai2007apparent,sheykhi2007thermodynamics,sheykhi2010entropy,cai2009entropic}.\\
	
To implement this, one follows the procedure in \cite{Sebastiani:2023brr} and identifies the total energy contained within the horizon as
	\begin{equation}
		E=\rho V=\rho\left(\tfrac{4\pi}{3}R_A^3\right).
	\end{equation}
Taking the differential,
	\begin{equation}
		\begin{split}
			dE&=\rho\,dV+V\,d\rho\\
			&=\left[\frac{4\pi}{3}R_A^3\, d\rho+2\pi R_A^2(\rho+p)dR_A\right]
			+W\,dV,
		\end{split}
	\end{equation}
comparing with the thermodynamic identity, we get
	\begin{equation} 
    \label{entropy}
	\begin{aligned}
        dS
		& = - (2\pi)^2 \left(1-\frac{\dot{R}_A}{2 H R_A}\right)^{-1}
		\left[ \tfrac{2}{3} R_A^4\, d\rho 
        \right. 
        \\
        & + \left. R_A^3(\rho + p)\, dR_A \right].
        \end{aligned}
	\end{equation}
Thus, for any given cosmological model, the entropy associated with the apparent horizon is obtained by integrating this expression, ensuring compatibility between the modified gravitational dynamics and the thermodynamic description.

\section{GBG model}
\label{sec:gbg}

The GBG model, introduced in 1977 \cite{regge2016general}, conceptualizes the universe as a four-dimensional brane embedded in a flat space of higher dimensions. In this proposal, the fundamental degrees of freedom are the embedding functions rather than the spacetime metric components. 
In particular, assuming that our four-dimensional universe is a hypersurface floating in 
a five-dimensional Minkowski space-time, the related dynamics is derived from the action
\begin{equation}
S[X^A,\psi]=\int d^4x\,\sqrt{-g}\left( \frac{R}{2\kappa_G}-L_{\text{m}}\right),\label{eq:action}
\end{equation}
where $X^A$ represents the embedding functions ($A = 0,1,2,3,4$), while $\psi$ denotes the matter fields defined on the brane and $R$ is the Ricci scalar defined on the world volume. Varying the action 
leads to the field equations 
\begin{equation}
\partial_\mu \left[ \sqrt{-g} \left( G^{\mu\nu} - \kappa_G T^{\mu\nu} \right) \partial_\nu X^A \right] = 0,   
\label{rt-eq0}
\end{equation}
where $G^{\mu\nu}$ is the Einstein tensor defined defined on the world volume, and $T_{\mu\nu}$ the energy-momentum tensor with ($\mu,\nu = 0,1,2,3$). In a more compact form we have a solely field equation given by
\begin{equation}
    \left(G^{\mu\nu} - \kappa_G T^{\mu\nu}\right)K_{\mu\nu}=0,
    \label{rt-eq1}
\end{equation}
where $K_{\mu\nu}$ the extrinsic curvature tensor. In our formulation, we will consider that the constituents of the Universe are characterized by a perfect fluid with energy density $\rho$ and pressure $p$. In passing, the expression \eqref{rt-eq1} is equivalent to  
\begin{equation}
    G_{\mu\nu}-\kappa_G T_{\mu\nu}=\tau_{\mu\nu}, 
    \label{rt-eq2}
\end{equation}
with condition $\tau_{\mu\nu}K^{\mu\nu}=0$, so the GBG equations can be seen as a generalization of 
the Einstein equations due to the presence of the additional term $\tau_{\mu\nu}$. In the cosmological framework, this extra geometrical contribution may be considered as an effective source of energy, indicating a potential connection with the dark energy paradigm \cite{davidson2001cold}. 

In a five-dimensional Minkowski space, one may embed an FLRW metric:  
\begin{equation}
    ds^2=-dt^2+\frac{a^2}{1-kr^2}\delta_{ij}dx^i dx^j,
\end{equation}
with the following choice of embedding functions
\begin{equation}
\begin{split}
    &X^0=\int^t \sqrt{1-\frac{\dot{a}^2}{k}}\, dt', \\
    &X^i=\frac{a x^i}{1+\tfrac{1}{4}kr^2}, \\
    &X^4=\frac{a}{\sqrt{k}}\left(\frac{1-\tfrac{1}{4}kr^2}{1+\tfrac{1}{4}kr^2}\right),
\end{split}
\label{embedding}
\end{equation}
in accordance with the local isometric embedding theorem in order 
to ensure a geometrically acceptable embedding~\cite{cartan1927, rosen1965}. Substituting~(\ref{embedding}) into~\eqref{rt-eq0} yields  
\begin{equation}
    \kappa_G \rho a^3\left(\dot{a}^2+k\right)^{1/2}
    -3a\left(\dot{a}^2+k\right)^{3/2}
    =-\frac{1}{9}\mu, \label{eq:master}
\end{equation}
where $\mu$ is an integration constant associated with the brane energy. This expression can be written as a cubic algebraic equation which, upon solving, leads to an effective Friedmann equation of the form  
\begin{equation}
    H^2+\frac{k}{a^2}=\frac{\kappa_G}{3}\,\xi \rho,
    \label{eff-eq}
\end{equation}
where the parameter $\xi$ satisfies  
\begin{equation}
    \xi(\xi-1)^2=\frac{\mu^2}{27\kappa_G^3\rho^3 a^8}.
    \label{eqxi}
\end{equation}
It is clear that when $\mu=0$, the nontrivial solution to \eqref{eqxi} is $\xi=1$, which leads the effective equation \eqref{eff-eq} to the Friedmann equation in the Einstein cosmology. In this framework, the parameter $\mu$ measures the degree of deviation from the standard cosmology. As mentioned earlier, we will focus on the scenario where $\xi \simeq 1$, implying that the deviations from general relativity introduced by the GBG model will be assumed to be minimal. Clearly, it is necessary to point out that the master 
equation~(\ref{eqxi}) admits solutions with distinct asymptotic behaviors. In particular, 
the vacuum case ($\rho \rightarrow 0$) leads to a regime where $\xi \rightarrow \infty$, indicating a significant deviation from general relativity; needless to say, 
other physical phenomena may appear in the remaining branch $0 < \xi < 1$.  Both regions, separated by the critical limit $\rho^3 a^8 \to \infty$, were discussed at length in~\cite{davidson1999cosmology,davidson2001cold}. As our focus in this work is on the 
minimal deviation regime ($\xi \simeq 1$), we restrict ourselves to matter 
contents where the perturbative expansion is valid, thus excluding the pure vacuum 
solution from the current thermodynamical analysis.

\section{Thermodynamics of GBG model}
\label{sec:thermo}

In this section, we examine the entropy corrections that occur in the GBG model, where the universe is described as an embedded brane. As demonstrated in the previous section, the dynamical equations are modified by the function $\xi$, which is linked to the energy of the brane. For simplicity and in agreement with the results of the current observations \cite{aghanim2020planck}, we focus on the scenario in which the spatial curvature is zero ($k=0$). Under these conditions, the apparent horizon plays the role of a causal boundary with radius $R_A=1/H$.

Starting from Eq.~\eqref{eff-eq}, taking the differential yields  
\begin{equation}
   d\rho = -\frac{6}{\kappa_G}\left(\xi'\rho+\xi\right)^{-1}\frac{dR_A}{R_A^3}.
   \label{drho}
\end{equation}
On the other hand, differentiating Eq.~\eqref{eff-eq} with respect to time and using the continuity equation $\dot{\rho}+3H(\rho+p)=0$, one obtains  
\begin{equation}
    \dot{H}=-\frac{\kappa_G}{2}\left( \xi' \rho+\xi\right)\left( \rho+p\right).
\end{equation}
Since the apparent horizon is the inverse of the Hubble parameter, this relation can be rewritten as  
\begin{equation}
    \rho+p=\frac{2}{\kappa_G}\left( \xi'\rho+\xi\right)^{-1}\frac{\dot{R}_A}{R_A^2}.
    \label{rhop}
\end{equation}
Substituting Eqs.~\eqref{drho} and \eqref{rhop} into the expression of entropy \eqref{entropy}, and rearranging the terms, leads to  
\begin{equation}
    dS_{H}=\frac{\left(4\pi\right)^2}{\kappa_G}\left(\xi'\rho+\xi\right)^{-1}R_A\, dR_A.
\end{equation}
It is clear that the entropy includes a correction represented by the factor $\left(\xi'\rho+\xi\right)^{-1}$. When the brane energy approaches zero, $\xi=1$, the formula simplifies to the conventional result found in general relativity. As usual, notice that the entropy increases by $dS_{H}$ as the radius of the horizon $R_{A}$ increases by $dR_{A}$. Our attention is now directed towards the scenario where the energy defined on the brane is small, ensuring that the effective equations remain nearly identical to the standard Friedmann equations. Under these conditions, the function $\xi$ can be developed in series around the nontrivial solution of Eq.~\eqref{eqxi}, resulting in the approximation
\begin{equation}
     \xi \simeq 1 + \frac{\beta^{1/2}}{\rho^{n/2}},
     \label{approxi}
\end{equation}
where $\beta=\mu^2/(3\kappa_G \rho_0^{8/3m})^3$ and we have considered the single fluid description and used that the energy density scales as $\rho=\rho_0/a^m$, allowing us to write $a=(\rho_{0}/\rho)^{1/m}$, with $m =3(1+\omega)$ being $\omega$ the constant parameter state identifying the cosmic component under consideration: $m=3$ ($\omega=0$) for dust and  $m=4$ ($\omega=1/3$) for radiation, for example. In addition, the exponent $n$ is determined by the relation $n=3-(8/m)=[9(1+\omega)-8]/[3(1+\omega)]$. This approximation is valid only for sufficiently small values of $\mu$, which guarantees that the adjustments introduced remain consistent. In this approximation, the energy density can be expressed as  
\begin{equation}
   \rho = \frac{ \beta^{1/n}}{(\xi-1)^{2/n}}.
   \label{rho2}
\end{equation}
Substituting Eq.~\eqref{rho2} into the effective Friedmann equation \eqref{eff-eq}, and considering the case of vanishing spatial curvature ($k=0$) where the Hubble parameter is simply the inverse of the apparent horizon radius, one obtains  
\begin{equation}
     \frac{1}{R_A^2} = \frac{\kappa_G\beta^{1/n}}{3}\,\frac{\xi}{(\xi-1)^{2/n}}.\label{eq:radius}
\end{equation}
Note that the definition of $\beta$ involves the parameter $\mu$, therefore the geometry of the apparent horizon can be directly related to the geometrical correction induced by the brane scenario, the parameter $\mu$. If $\mu$ is sufficiently small such that  
\begin{equation}
R_A^{2n}\kappa_G^n\beta/3^n \ll 1, \label{eq:condition}
\end{equation}
the function $\xi$ can be approximated as  
\begin{equation}
    \xi \simeq 1+ \left(\frac{\kappa_G}{3}\right)^{n/2}\beta^{1/2}R_A^n.
    \label{xiR}
\end{equation}
On the other hand, recalling the relation \eqref{approxi},  
by differentiating it with respect to \(\rho\), one finds 
\[
\begin{split}
\xi + \xi'\rho &= 1 + \frac{\beta^{1/2}}{\rho^{n/2}} - \frac{n}{2}\frac{\beta^{1/2}}{\rho^{n/2}} = \frac{n}{2} + \left(1 - \frac{n}{2}\right)\xi \\[6pt]
&= \frac{n}{2} + \left(1 - \frac{n}{2}\right)\left(1 + \left(\frac{\kappa_G}{3}\right)^{n/2}\beta^{1/2}R_A^n\right) \\[6pt]
&= 1 + \left(1 - \frac{n}{2}\right)\left(\frac{\kappa_G}{3}\right)^{n/2}\beta^{1/2}R_A^n,
\end{split}
\]  
where the approximation \eqref{xiR} was used. Substituting this result into the entropy expression \eqref{entropy}, we obtain  
\[
\begin{split}
dS_H &= \frac{(4\pi)^2}{\kappa_G}\,\frac{R_A\, dR_A}{1 + \left(1 - \tfrac{n}{2}\right)(\kappa_G/3)^{n/2}\beta^{1/2}R_A^n} \\[6pt]
&\simeq \frac{(4\pi)^2}{\kappa_G}\left[ 1 + \left(\tfrac{n}{2} - 1\right)\left(\frac{\kappa_G}{3}\right)^{n/2}\beta^{1/2}R_A^n \right] R_A\, dR_A.
\end{split}
\]  
In the second line, the denominator has been expanded under the assumption that the correction is small according to (\ref{eq:condition}). Notice that for $\beta=0$ the differential of the entropy becomes the standard one, $dS_{H} \propto R_{A}dR_{A}$. Finally, by integrating and reformulating the entropy in relation to the area of the apparent horizon, \(A = 4\pi R_A^2\), it becomes evident that at lowest order we can write
\begin{equation}
S_{H} = \frac{A}{4G}+\gamma \left(\frac{A}{4G}\right)^{\delta} + \mathcal{O}(1), 
\label{SH}
\end{equation} 
where for simplicity in the notation we have introduced the constants $\gamma, \delta$, defined as $\gamma \equiv \beta^{1/2}\left( \frac{3\omega - 5}{15\omega + 7}\right)\left(\frac{4G}{\sqrt{3}}\right)^{\tfrac{9\omega + 1}{3(1+\omega)}}$ and $\delta \equiv \tfrac{15\omega + 7}{6(1+\omega)}$. Notice that the first term represents the Bekenstein-Hawking entropy ($S_{BH}$) given by the quarter area law, while the correction terms that follow are expressed as a power-law of $S_{BH}$. We highlight that $S_{H}$ approaches $S_{BH}$, as $\beta$ tends to 0, confirming that our approach recovers the standard case in the appropriate limit.
On the other hand, we would like to highlight the fact that the correction term for the 
entropy is explicitly dependent on the value of $\omega$, consequently, entropy evolves differently for the various components that may be present in cosmic evolution. By way of illustration, when considering matter (\(\omega = 0\)), the entropy~(\ref{SH}) becomes
\begin{equation}
S_H = \frac{A}{4G} - \beta^{1/2}\frac{5}{7}\left(\frac{4G}{\sqrt{3}}\right)^{1/3}\left(\frac{A}{4G}\right)^{7/6},
\end{equation}
where a decrease in value is observed compared to the standard value.
These reductions frequently occur in entropy computations associated with modified 
gravity theories.

On the other hand, it is also possible to compute the temperature of the apparent horizon—or at least a good approximation of it—by assuming that \(\beta\) is very small. Using the approximation \eqref{approxi}, together with Eqs.~\eqref{superfi} and \eqref{eq:Tkappa}, one obtains the following result \footnote{Alternatively, for practical purposes, we write Eq. (\ref{approxi}) as
\begin{equation*}
    \xi = 1 + \lambda,
\end{equation*}
with $\lambda=\beta^{1/2}\rho^{-\frac{9\omega+1}{6(1+\omega)}}=\beta^{1/2}\rho_{0}^{-\frac{9\omega+1}{6(1+\omega)}}(1+z)^{-\frac{9\omega+1}{2}}$, which satisfies $\lambda \ll 1$ for $z\geq 0$ with $\beta \ll 1$.}  
\begin{equation}
T_A \simeq \frac{1}{4}\left(\frac{\kappa \rho_0}{3}\right)^{1/2}(1+z)^{\tfrac{3(1+\omega)}{2}}\left|(6\lambda+3)\omega - 1\right|.
\label{ta}
\end{equation}
Based on the approximations applied, it is assumed that \(\lambda \ll 1\). This prerequisite is crucial to ensure that the expansion remains consistent. Analogously to the entropy case, setting $\beta=0 \ (\lambda=0)$, reduces \eqref{ta} to the well-known expression for the apparent horizon temperature \cite{quevedo}.

\subsection{Bulk temperature and heat capacities}

\begin{center}
\begin{figure}[htbp]
  \centering
  \includegraphics[width=0.66\textwidth]{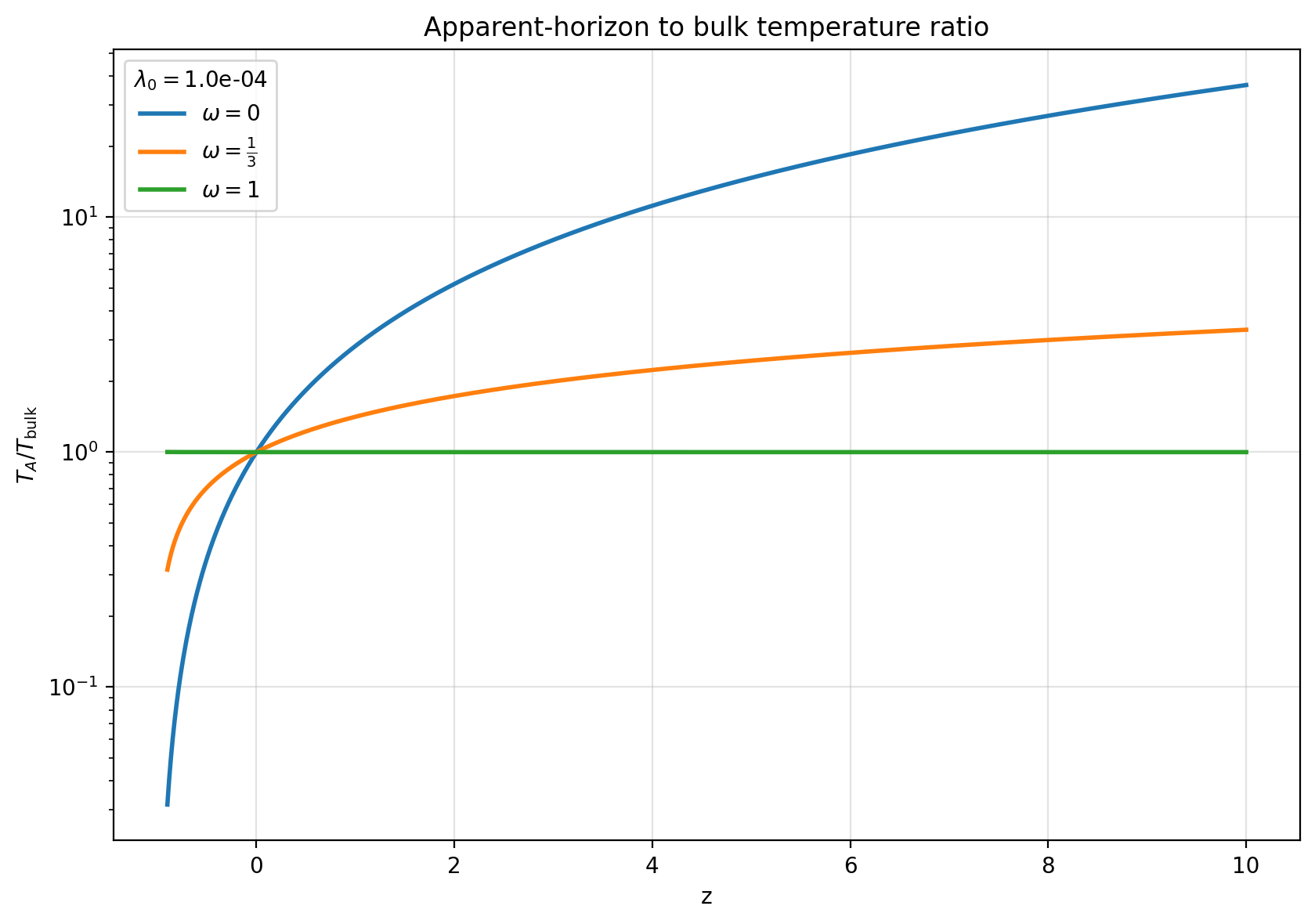}
  \caption{Ratio between the apparent horizon temperature and the bulk temperature for different types of matter.}
  \label{fig:mi-imagen}
\end{figure}
\end{center}

In the previous section, we calculated the corrections to entropy resulting from the universe's embedding and examined the impact on the apparent horizon's temperature. In this section, our focus shifts to determining the bulk temperature. For this purpose, we will consider the effective Friedmann equation \eqref{eff-eq}, which can be expressed as
\begin{equation}
    3H^2 = \frac{\kappa_G}{3}\bar{\rho},
\end{equation}
where we have assumed zero spatial curvature and defined the effective density as $\bar{\rho} = \xi \rho$. Recalling that matter content on the brane satisfies the continuity equation
\begin{equation}
   \dot{\rho} + 3H\rho(1+\omega) = 0,
   \label{or-con-eq}
\end{equation}
and by substituting $\rho = \bar{\rho}/\xi$ into the above equation, we obtain the following continuity equation for the effective density:
\begin{equation}
    \dot{\bar{\rho}} + 3H\bar{\rho}(1+\omega_{\text{eff}}) = 0,
\end{equation}
where an effective parameter state is identified and codifies the brane contribution as follows
\begin{equation}
\omega_{\mathrm{eff}} = \omega - \frac{\dot{\xi}}{3H\xi}.
\end{equation}
Since energy is conserved and the number of particles in the matter sector remains constant, we can adopt the line of reasoning given in the approach \cite{maartens1996causal, cardenas2019dark}, the bulk temperature can be linked to the effective equation-of-state parameter via
\begin{equation}
    \frac{\dot{T}_{\text{bulk}}}{T_{\text{bulk}}} = -3H\omega_{\text{eff}}.
\end{equation}
By substituting cosmic time with cosmological redshift using the standard relation $1+z=a^{-1}$ in the latter result, we integrate the resulting expression, allowing us to express it as
\begin{equation}
    T_{\text{bulk}} = T_0 (1+z)^{3\omega}\frac{\xi(z)}{\xi_0},
\end{equation}
where $T_0$ is the current time value ($z=0$) of the bulk temperature and $\xi_0 = \xi(z=0)$. If $\beta$ is sufficiently small, the above expression can be approximated as
\begin{equation}
    T_{\text{bulk}} \simeq T_0 (1+z)^{3\omega}\left( 1 + \lambda - \lambda_0 \right),
\end{equation}
with $\lambda_0 = \lambda(z=0)$. As can be seen, for $\omega=0$ the temperature exhibits small deviations from the constant value $T_{0}$, as obtained in the standard formulation, due to the presence of the term $\lambda$. The GBG model thus alters the thermodynamic description of the matter sector because the energy of the brane is responsible for the emergence of deviations from Einstein's theory. Assuming in addition that at the present epoch the bulk temperature coincides with the apparent horizon temperature, one finds that
\begin{equation}
    T_0 = \frac{1}{4}\left(\frac{\kappa_G\rho_0}{3}\right)^{1/2}\left|(6\lambda_0+3)\omega-1\right|.
\end{equation}
In view of this, we turn to 
analyze the ratio between both temperatures:
\begin{equation}
    \frac{T_A}{T_{\text{bulk}}} = (1+z)^{\tfrac{3}{2}(1-\omega)}
    \frac{\left|(6\lambda+3)\omega-1\right|}{\left|(6\lambda_0+3)\omega-1\right|(1+\lambda-\lambda_0)}.
\end{equation}
By plotting this ratio, 
from the Figure~\ref{fig:mi-imagen}, it can be seen that neither in the case of 
radiation, nor in the case of matter, does thermalization hold after $z=0$. 
The only component that thermalizes (and that in fact was already thermalized 
beforehand) is the \emph{stiff matter}~\cite{zeldovich1972hypothesis}. In this 
model, and under the assumptions adopted, we see that for matter and radiation 
the horizon is hotter than the bulk in the past, whereas in the future, the situation 
is reversed. Consequently, there is no persistence of thermal equilibrium between 
the apparent horizon and the bulk.
\onecolumngrid
\twocolumngrid

\begin{center}
\begin{figure}[h!]
    \centering
    \includegraphics[width=0.66\textwidth]{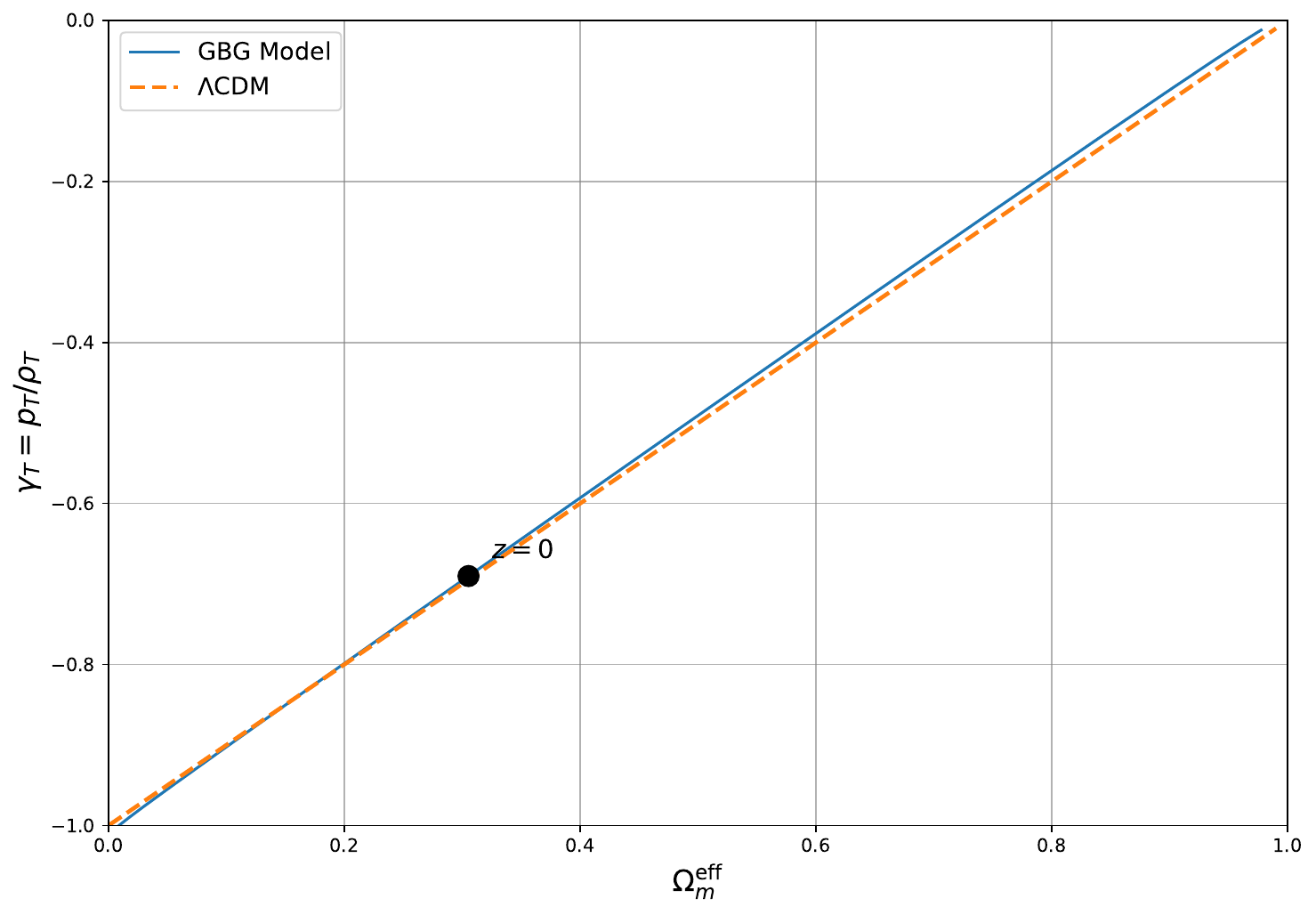}
    \caption{Evolution of the total equation of state as function of the cosmological redshift. We have considered the Planck 2018 results for $\Omega_{m0}$ \cite{aghanim2020planck}.}
    \label{fig:gamma}
\end{figure}
\end{center}

Employing conventional definitions of classical thermodynamics, we can write \cite{callen}
\begin{equation}
    C_{V} = \frac{\partial U}{\partial T}, \qquad C_{p} = \frac{\partial h}{\partial T}   
\end{equation}
Both expressions denote specific heat at constant volume $V$ and constant pressure $p$, respectively. Here, $U$ represents the internal energy of the system, while $h$ represents its enthalpy. $T$ is the temperature of the fluid. If we consider the volume enclosed by the apparent horizon, then we can write $U = \rho V = (4\pi/3) R^{3}_{A}\rho$ and $h=(\rho + p)V=(4\pi/3)(\rho + p)R^{3}_{A}$. Therefore, since all quantities depend on cosmic time, $C_{V}=(dU/dz)(dT/dz)^{-1}$ and $C_{p}=(dh/dz)(dT/dz)^{-1}$. If we focus on the case $\omega=0$, then the internal energy and enthalpy coincide under the assumption of a barotropic equation of state, leading to the same expressions for the specific heats, then we have the following from the use of Eqs (\ref{approxi}), (\ref{rho2}) and (\ref{eq:radius}) with $\kappa_{G}=1$
\begin{equation}
    C_{V,p}=-\frac{6\pi \sqrt{3}\rho^{5/12}_{0}}{T_{0}\beta^{11/4}}(1+z)^{5/4},
\end{equation}
which are negative and correspond to the lowest order contribution of the parameter $\beta$. Far from being an instability in the conventional fluid sense, negative heat capacity is a hallmark of self-gravitating systems, as established by Lynden-Bell \cite{Lynden_Bell_1999}. Just as a black hole heats up as it radiates (loses mass), the GBG model exhibits this inverse temperature-energy relationship. This confirms that the geometric corrections preserve the essential gravitational thermodynamic character of the apparent horizon, distinguishing it from a standard extensive gas. The fact that $\left|C_{V,p}\right| \gg 1$ arises from the parameter $\beta$, which satisfies the condition $\beta \ll 1$. 
Given that both specific heats have the same sign, we have an expanding accelerated universe as established in \cite{Saha_2025}. Since we assume that the matter content consists solely of the standard dark matter sector ($\omega = 0$), it implies that the geometric modifications introduced by the GBG model are sufficient to induce a state of accelerated cosmic expansion.
\onecolumngrid
\begin{center}
\begin{figure}[h!]
    \centering
    \includegraphics[width=0.66\textwidth]{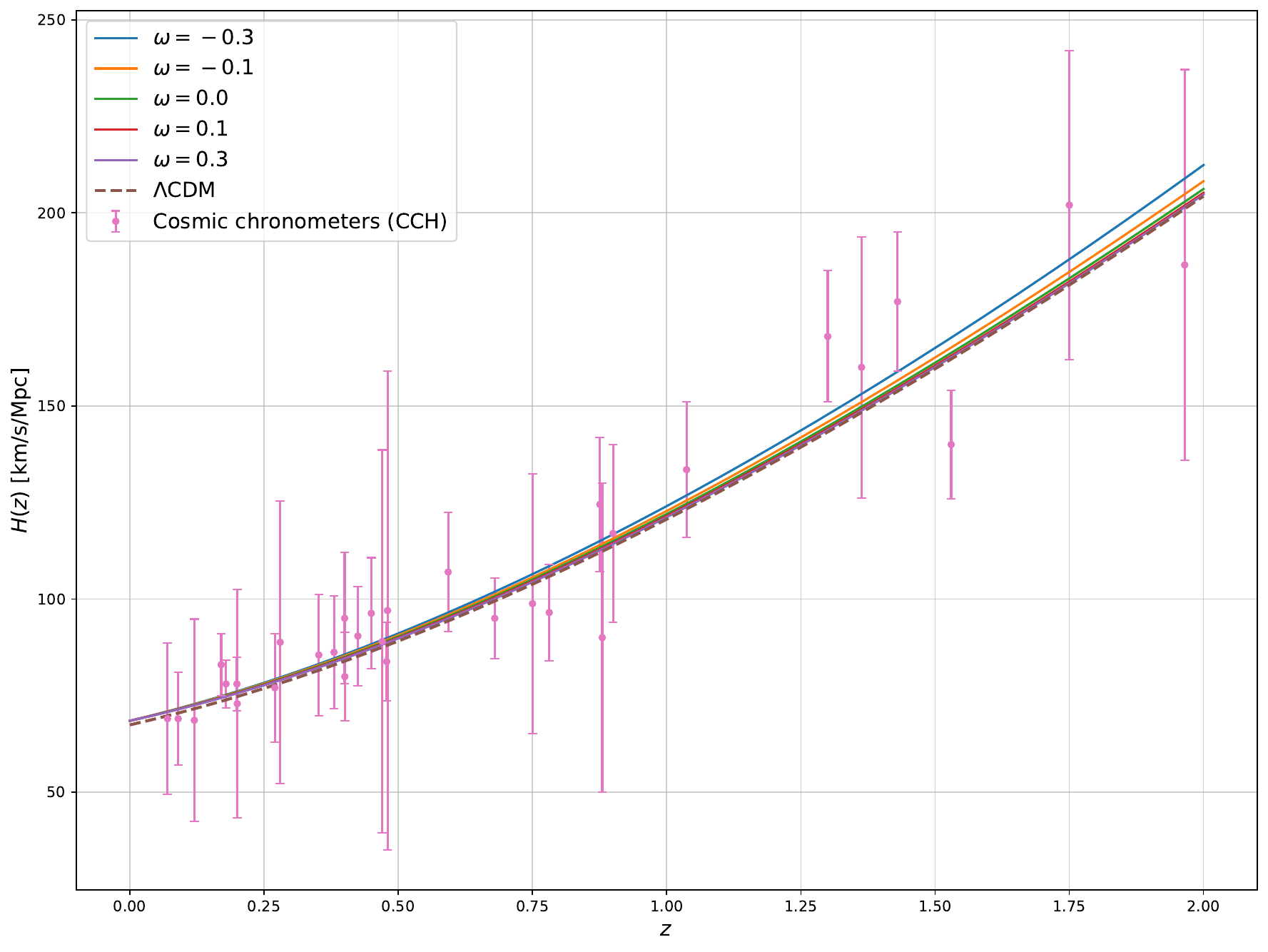}
    \caption{Hubble parameter as function of the cosmological redshift. We have considered the Planck 2018 results for the values of the cosmological parameters involved $(\Omega_{m0}, H_{0})$ \cite{aghanim2020planck}.}
    \label{fig:HubbleP}
\end{figure}
\end{center}
\twocolumngrid
To end our discussion, we perform a comparison of the GBG model at lowest order approximation against the standard $\Lambda$CDM model and the latest compilation of Hubble parameter measurements, which are given as function of the cosmological redshift obtained with cosmic chronometers \cite{di2023hubble}. In Fig. (\ref{fig:HubbleP}) we show the Hubble parameter obtained from (\ref{eff-eq}) with $k=0$, $\kappa_{G}=1$ and $\xi = 1+\lambda$, as discussed previously. In terms of energy density parameters, the Friedmann equation (\ref{eff-eq}) takes the form $H(z) =H_{0}\left[\xi(z) \sum_{i}\Omega_{i0}\,(1+z)^{3(1+\omega_{i})}\right]^{\tfrac{1}{2}}$. For a matter$+$ cosmological constant universe ($\omega_{m}=0$, $\omega_{\Lambda}=-1$), we have the following explicit Hubble parameter, $H(z) = H_{0}\left[\xi(z)\left\lbrace \Omega_{m0}(1+z)^{3}+\Omega_{\Lambda}\right\rbrace \right]^{\tfrac{1}{2}}$. Taking into account $\beta \approx\mathcal{O}(10^{-3})$, we observe that the correction induced by the GBG model under the approximation carried out in our analysis exhibits small deviations at an early time from the concordance model, while at late times they are practically indistinguishable; then the predictions obtained from the GBG model under this approximation remain close to the standard model, as expected. Consistent with the previous findings, the total equation of state of the GBG model obtained in our approach, which is defined as $\gamma_{T}\equiv p_{T}/\rho_{T}=-1+\frac{1+z}{3}\frac{d}{dz}\ln H^2(z)$ (where $H(z)$ is given as above), stays close to the $\Lambda$CDM predictions, as illustrated in Fig. (\ref{fig:gamma}); comparable results were reported in Ref. \cite{davidson2001cold}. Note that at the present epoch $(z=0)$, the universe displays a quintessence-like behavior associated with an accelerated expansion and asymptotically approaches a de Sitter phase.
\onecolumngrid
\twocolumngrid

\subsection{The cosmological constant case}
The perturbative $n$-analysis performed in arriving to
Eq. (\ref{approxi}) relies on the scaling $\rho \propto a^{-m}$ with 
$m\neq 0$. 
The scenario of a cosmological constant ($\omega = -1$) 
warrants a separate treatment, as it implies a constant density 
$\rho=\Lambda/\kappa_G$. Certainly, returning to the master 
equation (\ref{eqxi}) and assuming minimal deviations ($\xi=1+\epsilon$ with 
$\epsilon \ll 1$), we find the following
\begin{equation}
    \epsilon^{2} \simeq \frac{\mu^2}{27\Lambda^3 a^8} \quad \longrightarrow 
    \quad \epsilon \propto a^{-4}.
\end{equation}
This implies that the embedding geometry induces a correction term in the Friedmann equation that scales as $a^{-4}$. This mimics a {\it dark radiation} component, a feature previously identified in unified brane gravity scenarios \cite{Gurwich_2009}. Incorporating this radiative correction, the horizon temperature at early times is given by
\begin{equation}
    T_H=\left\vert\frac{\kappa_G}{6\pi}\left[ \rho_{dr}(1+z)^4-\frac{\Lambda}{\kappa_G}
    \right] \right\vert,
\end{equation}
where $\rho_{dr}=\kappa_G^{-1}\epsilon \Lambda a^4=\text{constant}$. While, conversely, the bulk temperature reads
\begin{equation}
    T_{\text{bulk}} = \frac{T_0}{\xi_0}(1+z)^{-3}\left[ 1+\frac{\kappa_G}{\Lambda}\rho_{dr}(1+z) \right].
\end{equation}
\noindent It is worth noting that, in this early epoch, the discrepancy between these expressions indicates that the universe is not yet in thermal equilibrium. Nevertheless, this correction dominates at small-scale factors but rapidly decays, leaving the standard de Sitter geometry as the stable late-time attractor. 
On thermodynamical grounds, this suggests that the embedding introduces an effective radiation entropy contribution at early times, which thermalizes with the bulk geometry as the universe expands.

\section{Final Remarks}
\label{sec:final}

In this work, we performed a thermodynamic analysis for the GBG model which extends the general relativity description by considering our universe as the result of the geodesical evolution of an extended object in an appropriate background. Our results indicate that within the context of geodetic brane gravity, the resulting scenario for the cosmological description of the model exhibits some relevant characteristics that could be useful in distinguishing it from the standard cosmological scenario. An important observation is that the entropy of the apparent horizon differs from the quarter area law because the model's dynamic equations differ from the standard Friedmann equations. When considering only small contributions, the lowest order corrections to the Bekenstein-Hawking entropy are given as power-law terms of the area, so $S_{H}=f(S_{BH})$, as usually obtained beyond general relativity. This demonstrates that connection between the universe's matter content and the geometric-thermodynamic properties of the horizon prevails, as already known.

The study of temperature evolution reveals that the thermal equilibrium between the apparent horizon and the bulk is not consistently maintained. Equilibrium for both matter and radiation is achieved only at present, and as the universe evolves, the components cool down due to cosmic expansion. A particularly interesting aspect is that for stiff matter a sustained thermalization between the horizon and the bulk is achieved. This result highlights the influence of the equation-of-state parameter on the thermal stability of the system and suggests that certain extreme matter regimes could favor a closer connection between cosmological dynamics and horizon thermodynamics.  

The analysis here achieved can be extended to scenarios with other interesting geometric structures.
In particular, the Lagrangian density presented in (\ref{eq:action}) can be extended by incorporating the invariant $K$, the trace of the extrinsic curvature of the world volume, as a gravitational term. This term characterizes the bending modes of the brane in the background spacetime. 
In this wise, the model given by $\mathcal{L} = R/(2\kappa G) + \zeta K$ represents a minimal extension of the GBG model that maintains the second-order nature of the equation of motion, and exhibits a self-accelerated branch in the cosmological scenario, which is interesting since it resembles the DGP braneworld results at some limit \cite{Cordero_2012}, in addition to being able to explore dark energy effects from this extended objects framework, \cite{Rojas2024dark}. For this extension to the model, equation (\ref{eq:master}) is modified by the appearance of a term of the form $(\zeta/a^{2})(\dot{a}^{2}+k)$, generalizing the solution obtained from the master equation (\ref{eqxi}), and providing more structure to explore. Nevertheless, these terms are not the sole options for describing brane cosmology. In fact, they serve as the initial terms in a series that can be developed by incorporating higher curvature terms, constructed through appropriate products of the extrinsic curvature of the world volume, by considering the repeated use of the contracted Gauss-Codazzi integrability condition. The result of this process is the analog of Lovelock gravity within the extended objects framework, and is referred to as {\it Lovelock type brane gravity} \cite{Cruz_2013}. We leave this topic open for further exploration.  Although we have highlighted that small contributions of the energy brane on the dynamical equations through the $\beta$ parameter are enough to induce an accelerating universe, no direct comparison with observational data has been made here. Statistical analysis using SNe Ia, BAO and CMB data to constrain $\beta$ and assess the fit quality remains a task for future work. Observational tensions such as the Hubble constant discrepancy deserve further exploration in this kind of cosmological scenario.\\ 

Overall, the results obtained indicate that geodetic brane gravity provides a consistent framework to explore how the thermodynamical properties of the universe may be altered by the presence of extra dimensions and effective energy parameters, offering valuable elements to better understand the interplay between geometry, matter, and cosmic thermodynamics.

\begin{acknowledgments}
 GC and GAP acknowledge SECIHTI support through the program {\it Estancias Posdoctorales por M\'exico 2023(1)}. This work was partially supported by S.N.I.I. (SECIHTI-M\'exico).
MC and ER acknowledge encouragement from ProDeP-M\'exico, CA-UV-320: \'Algebra, Geometr\'\i a
y Gravitaci\'on
\end{acknowledgments}

\bibliography{apssamp}
\end{document}